\documentclass{elsart}
\usepackage{natbib}
\usepackage{graphicx}
\begin{document}
\runauthor{Ramos, Wuensche, Ribeiro and Rosa}
\begin{frontmatter}
\title{Multiscaling and Nonextensivity of Large-scale Structures in the Universe}
\author[LAC]{F.M. Ramos,}
\author[DAS]{C.A. Wuensche,}
\author[IMECC]{A.L.B. Ribeiro,}
\author[LAC]{R.R. Rosa}

\address[LAC]{Laborat\'orio Associado de Computa\c c\~ao e Matem\'atica Aplicada, INPE,
12201-970, S\~ao Jos\'e dos Campos - SP, Brazil}
\address[DAS]{Divis\~ao de Astrof\'{\i}sica, INPE,12201-970, S\~ao Jos\'e dos Campos - SP, 
Brazil}
\address[IMECC]{Departamento de Matem\'atica Aplicada - IMECC, UNICAMP, 13083-970,
Campinas - SP, Brazil}
\thanks[]{This work was supported by FAPESP. FMR also acknowledges the support
given by CNPq through the research grant 300171/97-8.}
\begin{abstract}
There has been a trend in the past decade to describe the large-scale
structures in the Universe as a (multi)fractal set. However, one of the main 
objections
raised 
by the opponents of this approach deals with the
transition to homogeneity. Moreover, they claim there is not enough sampling space
to determine a scaling index which characterizes
a (multi)fractal set. In this work we propose an alternative solution to
this problem, using the generalized thermostatistics formalism.
We show that applying the idea of nonextensivity, intrinsic to this
approach, it is possible to derive an expression for the correlation
function, describing the scaling properties of large-scale structures
in the Universe and the transition to homogeneity, which is in
good agreement with observational data.
\end{abstract}
\begin{keyword}
cosmology, large-scale structures, fractal Universe,
nonextensivy, multiscaling, multifractals, generalized thermostatistics 
\end{keyword}
\end{frontmatter}

\section{Introduction}

One of the key problems in structure formation theories is the issue of the
galaxy distribution and the transition to homogeneity and isotropy of the 
Universe in large enough scales.
Historically, the hypothesis of homogeneity (the Cosmological Principle, CP) was
introduced by Einstein to find simple solutions of the field
equations for the case where the spatial hypersphere of the Universe is
a maximally symmetric subspace of the space-time.
That allows us to derive the Robertson-Walker metric and the
Friedmann equations, the theoretical framework where cosmology has been developed.
At the same time, the CP implies that all mass units should be 
statistically equivalent, with a Poisson distribution in space.
In this case, correlations may appear only on average and should be the same
when viewed from any point in the system. Actually, it has been
shown that luminous matter in the Universe shows a quite structured 
distribution of galaxies and voids. 
This structure has led some authors [1] to tentatively describe galaxy
clustering  using a fractal distribution with a dimension $\sim 1.2$.
The results of a number of redshift surveys [2,3] indicate that, indeed,
there is a certain hierarchy in the Universe, with stars forming galaxies,
galaxies grouping themselves in groups, clusters and superclusters.
A rough limit to this clustering is seen at a distance of
about 200 $h^{-1}$ Mpc and the present redshift surveys do not show evidence
of large-scale structures beyond this scale.
Particularly, in very large scales - those probed by cosmic microwave background (CMB)
experiments, especially by the COBE-DMR experiment -
there is no evidence of violation of local isotropy [4].
However, we should note that the existence of such a crossover towards homogenization, as well
as the exact value of the fractal dimension, are questions of intense debate [5-7].

The fractal hypothesis is deeply connected with a topology theorem which states
that homogeneity is implied by the condition of local isotropy plus the assumption
of analicity (or regularity) for the distribution of matter. It is possible to
prove that in a fractal the condition of local isotropy can be satisfied
but, since a fractal is a non-analytic structure, the property of homogeneity is not
implied [8]. This means that the fractal scenario is incompatible with the CP. 
Observations point out that galaxy structures indeed exhibit
fractal properties up to some scale, although a pure fractal description does not
seem to be favored at the moment [9]. Recently, some authors (e.g. [10-12]) have
had some success in describing the clustering properties of visible matter in the
Universe in terms of a multifractal phenomenon associated with density thresholds
applied to multifractal sets. They show that both a hybrid fractal and a
multifractal approaches can reasonably describe the matter distribution up to
$\approx~100h^{-1}$ Mpc. 

In spite of the relative success of the multifractal approach to describe the distribution
of matter in the Universe, it is important to understand the physics behind this
framework. In general terms, the multifractal description of galaxies may
represent a strange attractor that is the nonlinear outcome of the
dynamical equations of gravitational galaxy clustering [13,14]. However, it is
not simple to find a dynamical connection between fractal sets  and
galaxy clustering. In this work we propose an alternative solution to
this problem, using the generalized thermostatistics (GTS) formalism [15].
We show that applying the idea of nonextensivity, intrinsic to GTS, 
it is possible to derive an expression for the correlation
function, describing the scaling properties of large-scale structures
in the Universe.  The present approach is based on the assumption of
a scale dependent correlation dimension ($D_2$), which leads to a reconciliation
with observational data at various scales, showing  a smooth transition
from a clustered, fractal Universe to large-scale homogeneity, with $D_2 = 3$.

The physical motivation behind our approach is the peculiar behavior of
large-scale gravitational systems, dominated by the unshielded, long-range nature of 
gravity. In contrast, other many-body systems, like neutral gases and plasmas, 
are characterized by short-range interactions. Because of this fundamental difference, the standard 
Boltzmann-Gibbs statistical mechanics cannot be applied to gravitating 
systems, since the long-range nature of gravity strongly violates one of its basic
premises (short-ranged effective interactions), and, thus, 
special techniques are needed [16]. One possibility to overcome this difficulty -- adopted 
by us -- is to use the GTS, a theory proposed to correct 
Boltzmann-Gibbs statistical mechanics in those cases where 
its prescriptions fail.

\section{Theoretical Framework}

In the context of galaxy clustering, the important exponent is the
correlation dimension $D_2$ [10], which is defined in terms of the scaling
of the correlation integral $C(r)$ over a distance $r$ (normalized 
by a reference scale L):

\begin{equation}
\label{corr1}
C(r) = \int_0^r 4 \pi n (1 + \xi(s))s^2 \, ds = A r^{D_2}~~,
\end{equation}

\noindent where $\xi(r)$ is the two-point correlation function, $n$ is the
mean density of objects and $A$ is a constant.
By differentiating eq. (\ref{corr1}) with respect to $r$ we get:

\begin{equation}
1+\xi(r) = \frac{D_2  \, A \, r^{(D_2-3)}}{4 \, \pi \, n }~~.
\end{equation}

Considering that in large enough scales we have $(1 + \xi) \rightarrow 1$ and 
$D_2 \rightarrow 3$, 
we obtain
$A = 4 \, \pi n / 3$ and

\begin{equation}
\label{corr2}
1+\xi(r) = \frac{D_2 \, r^{(D_2-3)}}{3}~~.
\end{equation}

Based on the scaling properties of multifractals, a generalization of Boltzmann-Gibbs
thermostatistics has been proposed [15] through the introduction of a family
of generalized entropy functionals $S_q$ with a single parameter $q$.
These functionals reduce to the classical,  extensive Boltzmann-Gibbs form
as $q \rightarrow 1$.
The generalized entropy has been successfully used to describe a wide range
of phenomena, including long-range interactions [17], 
turbulence [18-20], anomalous diffusion [21], among others. For an up-to-date list of references
on GTS and its applications, see [22].

For a system with
$W$ microscopic state probabilities $p_i\geq 0$, that are normalized
according to $\sum_i^W p_i = 1$, the GTS formalism is based upon two axioms. 
First, the entropy of the system is given by
\[
 S_q = k \frac{1-\sum_i^W p_i^q}{q-1}
     = \frac{k}{q-1}\sum_i^W \left( p_i - p_i^q \right),
\]
where $k$ and $q$ are real constants. 
Second, an experimental measurement of an observable $O$, 
whose value in state $i$ is $o_i$, yields the (unnormalized) 
$q$-expectation value,
\[
 O_q = \sum_i^W p_i^q o_i,
\]
of the observable $O$.

Within the framework of GTS, the following special additivity rule holds:
\begin{equation}
\label{ca1}
O_q(2V) = 2 O_q(V) + 2 (1-q) O_q(V) S_q(V)/k ~~,
\end{equation}
\noindent where $V$ is an arbitrary volume. In the sense of eq. (\ref{ca1}),
the entropic index $q$ characterizes the degree of nonextensivity of the system.
 
Nonextensivity implies that $O_q$ is not uniformly distributed within
$V$ but rather concentrated on a subset of noninteger dimension $D_2$.
In this case, the following scaling relation holds:
\begin{equation}
\label{ca2}
O_q(2V) = 2^{D_2/3} O_q(V)~~.
\end{equation}

The reasoning above may be extended to the context of galaxy clustering
if we assume the correlation integral $C(r)$ as being the observable in
eq. (\ref{ca1}).
Thus, with the additional assumption of a scale 
dependent entropic parameter $q(r)$,
from eqs. (\ref{ca1}) and (\ref{ca2}), 
it immediately follows:
\begin{equation}
\label{ca3}
D_2(r) =  3 \, \frac{\log(2 +  a (1 - q(r)))}{\log 2}~~,
\end{equation}
\noindent where $D_2(r)$ is the correlation dimension and $a=2 S_q/k$ 
will be determined later.

The idea of relating nonextensivity and multiscaling through 
a scale dependent correlation dimension, as given by eq. (\ref{ca3}), has been originally proposed by us
in the context of fully developed turbulence [18,19]. Interestingly, Chen and Bak [23] 
recently suggested
a similar relation, derived from a simple reaction-diffusion model 
of turbulent phenomena, as a new geometric form for describing the distribution of
luminous matter in the Universe [24].

To be complete, the above formulation shall contain a model for $q(r)$.
We suggest 
the following simple {\it ansatz}:
\begin{equation}
\label{ca4}
r \sim 1/(q-1)^{\beta}~~,
\end{equation}
\noindent or equivalently $(q-1) \sim (1/r)^{1/\beta}$, with $\beta > 0$.
At large scales, as $r$ grows, we have $q \rightarrow 1$ and $D_2 \rightarrow 3$. 
The appropriate choice
of 
$a$ sets the value of $D_2$ at a lower reference scale.

\section{Results and Discussion}

\begin{figure}[ttt]
\begin{center}
\includegraphics[width=10cm]{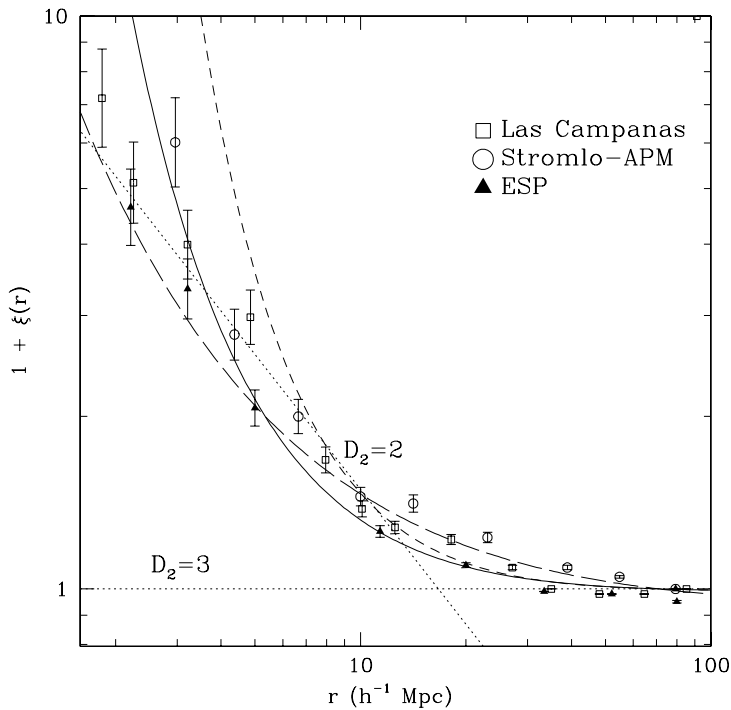}
\end{center}
\caption{The two-point correlation function versus scale,
for the Stromlo-APM, the Las Campanas and
the ESP redshift surveys [3], and for the present model, for $a=0.65$ and $\beta=1.0$
(solid line); $a=1.60$ and $\beta=0.8$
(dashed line); $a=0.28$ and $\beta=2.0$
(long dashed line); with $L=100 h^{-1}$ Mpc for all cases. For comparison purposes, 
a purely fractal description ($D_2=2$) and a 
homogeneous scenario ($D_2=3$) are also displayed (dotted straight lines). }
\end{figure}

We compared our model against observational data from various redshift surveys [2,3].
Results are shown in Figs. 1 and 2, for $a=0.65$ and $\beta=1.0$
(solid line); $a=1.60$ and $\beta=0.8$
(dashed line); $a=0.28$ and $\beta=2.0$
(long dashed line); with $L=100 h^{-1}$ Mpc for all cases.
For comparison purposes, a purely fractal description ($D_2=2$) and a 
homogeneous scenario ($D_2=3$) are also
depicted in Fig. 1 (dotted straight lines).
We observe that for $\beta = 1$, which corresponds to the simplest relation
between $q$ and $r$, our results show a good agreement with observational
data for both the two-point correlation function and the correlation
dimension. Increasing $\beta$, we may obtain a better fitting for
$1+\xi(r)$ at the expense of the correlation dimension data. The
opposite is true for decreasing values of $\beta$.

\begin{figure}[ttt]
\begin{center}
\includegraphics[width=10cm]{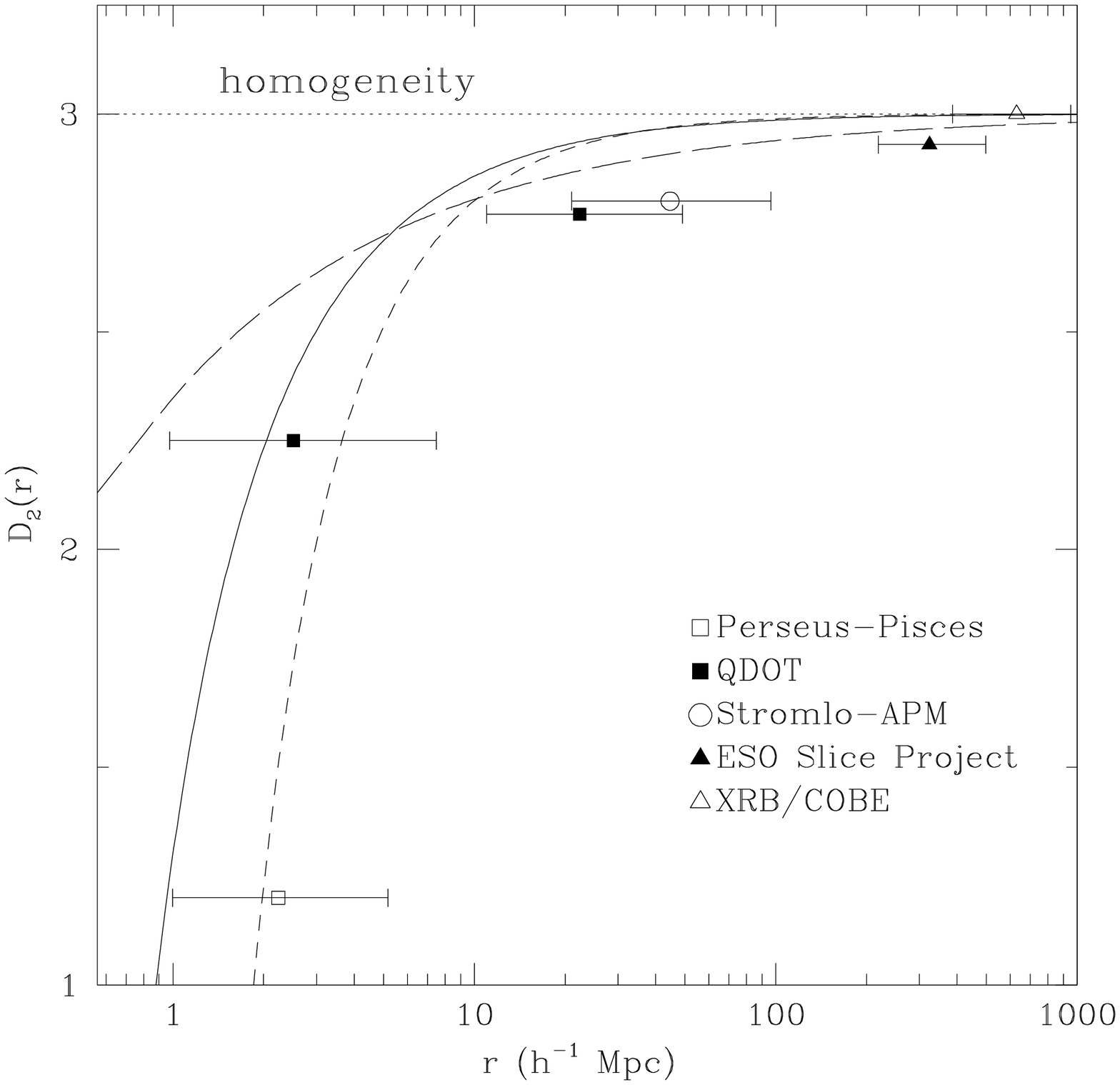}
\end{center}
\caption{The correlation dimension versus scale,
for various surveys [2], and for the present model, for $a=0.65$ and $\beta=1.0$
(solid line); $a=1.60$ and $\beta=0.8$
(dashed line); $a=0.28$ and $\beta=2.0$
(long dashed line); with $L=100 h^{-1}$ Mpc for all cases.}
\end{figure}

Equations (\ref{corr2}) and (\ref{ca3}) offer a quantitative description of matter clustering
in the Universe, and of the smooth transition from small-scale nonextensive
fractal behavior to large-scale extensive homogeneity. From a geometrical point of view, 
our  model shows a Universe displaying a clear hierarchy, with predominance 
of point-like ($D_2 \sim 0$) and filamentary ($D_2 \sim 1$) structures at small scales,
and surface-like ones ($D_2 \sim 2$) at intermediate scales. For sufficiently large scales 
($r > 500 h^{-1} Mpc$), the homogeneity predicted by the Cosmological Principle 
is recovered. All these features of the present approach are in good agreement with observational
data. 

Summarizing, we may say that our primary motivation in this work was to investigate 
the prediction
of multiscaling and nonextensivity of large-scale structures in the
Universe within the context of the generalized thermostatistics formalism.
The results presented above suggest that we cannot discard this 
theoretical
framework as a viable way to explain the gravitational
clustering in the Universe. However, we should be careful
and keep in mind two important caveats: the poorness
of observational data in some 
scale domains may be masking the
results, and the fact that 
luminous galaxies
may not perfectly trace the mass distribution (the biasing problem). 
From a theoretical point of view, what is missing at the moment is a detailed understanding 
of how $q$ vary with scale. In the present work, we adopted the simplest model that provided a good
agreement with the data. We are currently running a number of COBE-normalized CDM 
simulations for some relevant cosmological models [25]. We believe that further study on the 
resulting correlations and velocity distributions, at different scales 
(from galaxies to superclusters), may reveal the connection 
between the entropic parameter and the dynamics of mass clustering 
in the Universe. 

\ack{This work was partially supported by FAPESP and CNPq-Brazil.}

\end{document}